# High Photon Upconversion Efficiency with Hybrid Triplet Sensitizers by Ultrafast Hole-Routing in Electronic-Doped Nanocrystals

Alessandra Ronchi[1], Chiara Capitani[1,2], Graziella Gariano[2], Valerio Pinchetti[1], Matteo Luca Zaffalon[1], Francesco Meinardi[1,2], Sergio Brovelli[1,2*] and Angelo Monguzzi[1*]

[1]*Dipartimento di Scienza dei Materiali, Università degli Studi Milano Bicocca, via R. Cozzi 55, 20125 Milan, Italy*
[2]*Glass to Power SpA, Via Fortunato Zeni 8, I-38068 Rovereto, Italy*

**Low power photon upconversion (UC) based on sensitized triplet-triplet annihilation (*s*TTA) is considered as the most promising upwards wavelength-shifting technique to enhance the light harvesting capability of solar devices by recovering the low-energy tail of the solar spectrum. Semiconductor nanocrystals (NCs) with conjugated organic ligands have been proposed as broadband sensitizers for extending the light-harvesting capability of *s*TTA-UC molecular absorbers. Key to their functioning is efficient energy transfer (ET) from the NC to the triplet state of the ligands that sensitizes the triplet state of free emitters, whose annihilation generates the upconverted emission. To date, the triplet sensitization efficiency in such systems is limited by parasitic processes, such as charge transfer (typically of the photohole) to the organic ligand due to the disadvantageous band alignment of typical NCs and organic moieties. Available strategies only partially mitigate such losses and intrinsically limit the ET yield. Here we demonstrate a new exciton-manipulation approach that enables loss-free ET without detrimental side-effects. Specifically, we use CdSe NCs doped with gold atoms featuring a hole-accepting state in the NC bandgap at a higher energy than the HOMO level of the ligand 9-anthracene acid. Upon photoexcitation, the NC photoholes are routed to the Au-state faster than their transfer to the ligand, producing a long-lived bound exciton in perfect resonance with its triplet state. This hinders hole-transfer losses and results in ~100% efficient ET, over 50-fold higher than in standard NCs. By combining our hybrid sensitizers with an annihilator moiety, we achieved an *s*TTA-UC efficiency of ~12% (~24% in the *normalized* definition), which is the highest value for hybrid upconverters based on *s*TTA reported to date and approaches optimized organic systems.**

Triplet excitons are common in organic semiconductors, but generally inherently difficult to access because they involve spin-forbidden electronic transitions. Nevertheless, thanks to the recently improved ability to manage spin-flip processes,[1-4] triplet excitons have been receiving significantly increased attention for their potential application in solar technologies,[1] as in the case of photon upconversion (UC) systems based on sensitized triplet-triplet annihilation (*s*TTA).[5-7] By *s*TTA, upconverted photons are generated by the radiative decay of fluorescent singlet states obtained through the fusion after collision of optically dark triplets of two annihilator molecules (often referred to also as *emitters*). Such triplet states are populated via Dexter energy transfer (ET) from a low-energy light-harvesting moiety, commonly referred to as the *sensitizer*.[8-11] With optimized organic sensitizer/annihilator pairs UC quantum yields ($QY_{uc}$) as high as 30% have been obtained at excitation intensities comparable to solar irradiance, not far from energy-conservation limit of $QY_{uc} = 50\%$.[12] Crucially, the *s*TTA-UC process is activated upon absorption of non-coherent photons and its efficiency at solar irradiance levels is orders of magnitude higher than typical non-coherent light upconverters based on lanthanide ions.[13,14] For this reason, *s*TTA-UC is considered as the most promising upwards wavelength-shifting approach to recover the sub-bandgap portion of the solar spectrum of most photovoltaic or photocatalytic devices.[8,15-21]

Despite this promise, the application of *s*TTA-UC is still hindered by the limited choice of organic moieties with optically accessible long-lived triplet states (i.e. sufficiently efficient intersystem crossing) necessary to act as triplet sensitizers and by their typically narrow absorption bandwidth, which caps their spectral-harvesting capability of solar light. To overcome this issue, several strategies have been introduced, such as the use of multicomponent or multilayer systems exploiting sensitizers with complementary absorption properties.[18,20,22,23,24] More recently, colloidal semiconductor nanocrystals (NCs)[25,26] have been proposed as broadband-absorbing component of hybrid *s*TTA-UC sensitizers.[27-31] NCs offer several advantages over molecular photosensitizers, including facile preparative synthesis via scalable colloidal methods, photostability, size-tunable electronic and photophysical properties,[26] large molar extinction coefficients and broadband optical absorption extendable from the UV to the near-infrared spectral region,[32,33] which is key for coupling *s*TTA-UC with silicon-based devices. Furthermore, NCs feature a versatile surface chemistry that enables their post-synthesis functionalization with a variety of molecular species, including suitable derivatives of conjugated organic dyes.[34] As a result, as sketched in **Figure 1A** for the case of CdSe NCs functionalized with 9-anthracenecarboxylic acid (9-ACA), NCs and conjugated dyes can be rationally combined so as to create hybrid sensitizers[27-29,35] in which the NC absorbs and transfers the luminous energy via Dexter ET (ET′ in **Figure 1A**) to the triplet state of a surface-attached organic moiety (hereafter referred to as *triplet collector,* TC). The resulting excited triplet state acts as an energy bridge to populate, via a second Dexter-type ET process (ET′′), the long-lived triplet state of a free emitter, whose fusion with the excited triplet state of a second emitter (generated through an identical multi-step process) finally generates a highly-emissive singlet responsible for the upconverted luminescence.[6 8-10]

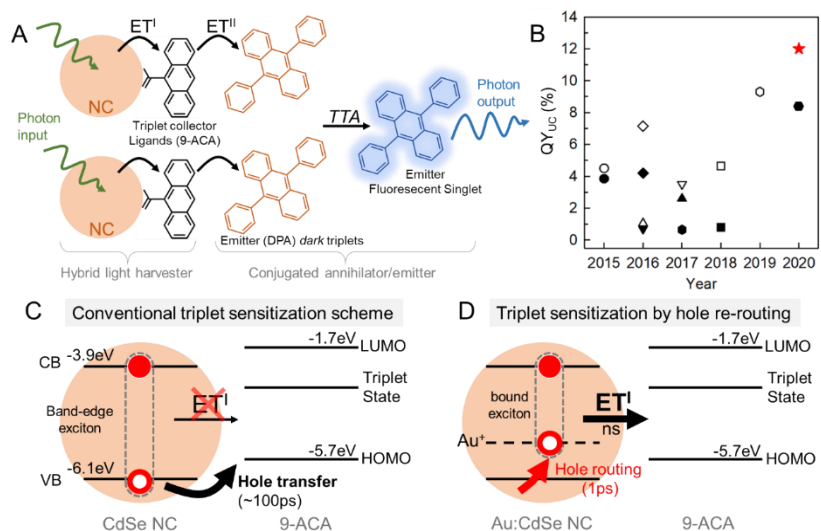

**Figure 1 | Sensitized *s*TTA-UC in hybrid NC-based systems and hole re-routing strategy using electronic-doped NCs. A.** Energy flux in *s*TTA-UC. Upon absorption of a green photon by the NC, the exciton energy is transferred via energy transfer (ET′) to the triplet collector ligand that then populates the triplet state of the emitter via ET′′. The annihilation of two emitter triplets (TTA) results in the formation of a high-energy fluorescent singlet state responsible for the upconverted blue luminescence. **B.** Chronological evolution of the *s*TTA-UC quantum yield $QY_{uc}$ in hybrid systems since their appearance in 2015.[28,29,33-43] The red star marks the efficiency obtained in this work. The $QY_{uc}$ values are reported according to the standard convention (Supplementary Material). **C**. Energy diagram and mechanistic picture of the hole-transfer process that outcompetes ET′ in CdSe NCs functionalized with 9-ACA. **D**. In CdSe NCs with Au$^+$ dopants, hole transfer is outpaced by the ultrafast localization of the photohole in the intragap states (1-2 ps vs. ~100 ps) associated with the *d*-levels of Au$^+$, leading to the formation of a bound exciton perfectly resonant with the triplet state of 9-ACA. As a result, the ET′ step reaches 100% efficiency.

Over the years, tremendous progress in the design of such NC-organic hybrid sensitizers has enabled a substantial growth of the *s*TTA-UC performance, as highlighted in **Figure 1B**, where we report the chronological evolution of the $QY_{uc}$ since their original appearance in 2015.[28,29,33-43] However, despite such advancements, hybrid *s*TTA-UC systems are still not as performing as the best fully-organic counterparts.[32,36-39] This is mainly due to parasitic processes that can limit the efficiency of the ET′ step that populates the TC triplets. For example, non-radiative trapping of band-edge (BE) photo-carriers in defect states on the NC surfaces, which might become particularly abundant when the native ligands used for the NCs synthesis are replaced with more bulky TCs, resulting in incomplete surface coverage [40]. Although this problem can be mitigated by shelling the NCs surfaces with wider energy gap semiconductors[41-43], other loss phenomena are more difficult to address, as they are inherently related to the electronic structure of typical NCs and TC molecules. Specifically, the highest occupied molecular orbital (HOMO) of most TCs lies above the valence band (VB) energy of most II-VI or IV-VI semiconductors, such as Zn, Cd or Pb chalcogenides[44] which, in NCs, is further shifted to lower energies with respect to vacuum by quantum confinement.[26] As a result, as shown in **Figure 1C** for the (CdSe NC)-(9-ACA) hybrid sensitizer, fast transfer (~ hundreds of ps) [40 45] of the photo-hole from the NC's VB to the HOMO of 9-ACA can result in efficient exciton dissociation that outcompetes ET′ and thereby hinders the upconversion process.[42,46] We highlight that, because of the heavier hole effective mass with respect to the electron in many chalcogenide semiconductors, the VB energy is less sensitive to quantum confinement with respect to the CB. As a result, in such systems size control is ineffective to prevent hole transfer, because the limited upward energy shift of the VB at increasing particle size does not allow to rise its energy above the TC HOMO. Moreover, the increasing size is accompanied by an even larger lowering of the CB energy, resulting in red-shifted BE exciton emission that can disrupts the spectral resonance with the triplet state of the TC and thereby diminishes the efficiency of ET′ ($\Phi_{ET′}$). On a similar line, wide energy gap shelling suppresses carrier transfer from core/shell NCs to surface ligands, but the increased donor-acceptor distance between the NC core and the TC lowers the ET′ rate in favour of radiative exciton decay, as shown for PbS/CdS and CdS/ZnS NCs functionalized with (5-carboxylic acid tetracene)[37,42] or 2,5-diphenyloxazole molecules.[41]

A strategy for suppressing efficiency losses associated to hole-transfer, that has not been explored so far, would be to employ NCs with an engineered intragap hole-accepting state lying above the HOMO energy of the TC and featuring a higher hole-capturing rate. For this scheme to be effective, upon photoexcitation, ultrafast re-routing of the BE hole into such an intragap state should generate a bound exciton - featuring a CB electron orbiting within the Coulomb potential of a localized photohole - with energy matching the triplet state of the TC. Such a design would provide multiple benefits for the upconversion process: the intragap photohole would be unaffected by transfer to the TC HOMO and, owing to the reduced spatial overlap between the electron and hole wavefunctions, the exciton radiative recombination rate would be significantly lowered with respect to the BE exciton, thus favouring Dexter ET′ over radiative decay.[29] Also crucially, by operating the energetics of the hole transfer process, this scheme does not require heterostructuring with thick wide-bandgap shells and would thus impose no limitation to ET′ due to increased donor-acceptor distance.

In this work, we realize this regime for the first time by exploiting the ultrafast hole dynamics in CdSe NCs electronically doped with $Au^+$ cations. In these systems, similar to $Cu^+$- or $Ag^+$-doped NCs [47-49], the $d$-electrons of $Au^+$ impurities (in $5d^{10}$ electronic configuration) introduce intragap hole acceptor states at $\Delta E_{VB-Au}$~0.6 eV above the VB maximum that localize the photohole in ~1-2 ps.[47] This gives rise to a bound exciton with the delocalized CB electron, whose radiative decay yields the characteristic long-lived, Stokes-shifted photoluminescence (PL). Importantly, since the $Au^+$ states are pinned to the host VB, the energy of the bound excitons can be tuned by control of the particle size so as to maximize the energy resonance with the triplet state of the TC moiety of choice, 9-ACA. This represents the best candidate for validating our hole-routing concept in an established $s$TTA-UC hybrid, having been used for demonstrating interfacial Dexter-like energy transfer from CdSe NCs to surface-anchored acceptors triplets[27] and, more recently, in high performance systems[29,35]. Crucially, as depicted in **Figure 1D**, $\Delta E_{VB-Au} > \Delta E_{VB-HOMO}$ (~0.2 eV) and the hole-localization time in the $Au^+$ center is much faster than typical hole extraction processes,[40,45] leading to complete suppression of hole-transfer losses in advantage of the ET′ ($\Phi_{ET'}$~100%). As a result, by combining our engineered (Au:CdSe NCs)-(9-ACA) sensitizers with 9,10-dyphenilanthracene (DPA) upconverting organic molecules in solution,[8] we obtained the unprecedented $QY_{uc}$~ 12% (~24% using the *normalized* definition of upconversion yield often found in the literature[25,27,31-40]. See Supplementary Material), which represents the highest performance for hybrid $s$TTA-based upconverters reported to date. Possibly more importantly, such an efficiency is approaching the best fully-organic $s$TTA-UC systems, thus potentially paving the way for the application of NC-based upconverters in real-world technologies.[10,50]

CdSe NCs doped with $Au^+$ impurities were synthesized following the method reported in ref.[47] using oleic acid (OA) as surface ligands. The particle size is 2.3 nm (Supplementary Figure S1) so as to maximize the spectral resonance between the emission energy of the doped NCs, pinned to the VB, and the triplet state of the TC. After purification, the native OA ligands were replaced by 9-ACA moieties. Details of the synthesis procedure and structural characterization of the NCs are reported as the Supplementary Material. To quantitatively evaluate the effectiveness of the hole routing strategy, we performed side-by-side spectroscopic investigations of toluene dispersions (with identical absorbance of 0.13 OD at 532 nm) of undoped and Au-doped CdSe NCs before and after ligand exchange with 9-ACA molecules. **Figure 2A** and **2B** show the optical absorption and PL spectra of CdSe and Au:CdSe NCs both as-synthesized (OA-capped) and functionalized with 9-ACA; the corresponding PL decay curves are shown in **Figure 2C** and **2D** respectively. Both doped and undoped NCs show a well-defined 1S absorption peak confirming the good size dispersion of the NCs and that the incorporation of $Au^+$ impurities does not perturb the electronic structure of the NC host, in agreement with previous studies[47]. The as-synthesized undoped NCs feature the typically narrow band-edge PL of CdSe NCs (Stokes shifted of ~80 meV from the respective 1S absorption peak) and a weak broadband emission at lower energies commonly ascribed to surface traps.[51] The time decay of the band-edge PL (**Figure 2C**) is multi-exponential consistent with the core-only nature of the NCs and features an effective lifetime $\tau^{OA}_{Band\ Edge}$~ 40 ns (Supplementary Table 1). Doping with $Au^+$ drastically modifies the PL spectrum and dynamics of the NCs, with the suppression of the band-edge PL in favor of a broad, long-lived emission centered at ~670 nm (1.85 eV). In agreement with the mechanistic scheme drawn in **Figure 1D** and in the inset of **Figure 2B**, as well as with previous reports on NCs doped with group 11 metals,[47-49] this intragap emission, hereafter referred to as Au-PL, is ascribed to the radiative decay of the CB electron in the $d$-states of gold following the transient oxidation of $Au^+$ to $Au^{2+}$ upon ultrafast localization of the VB hole according to the reaction $Au^+ + h_{VB} \rightarrow Au^{2+}$.

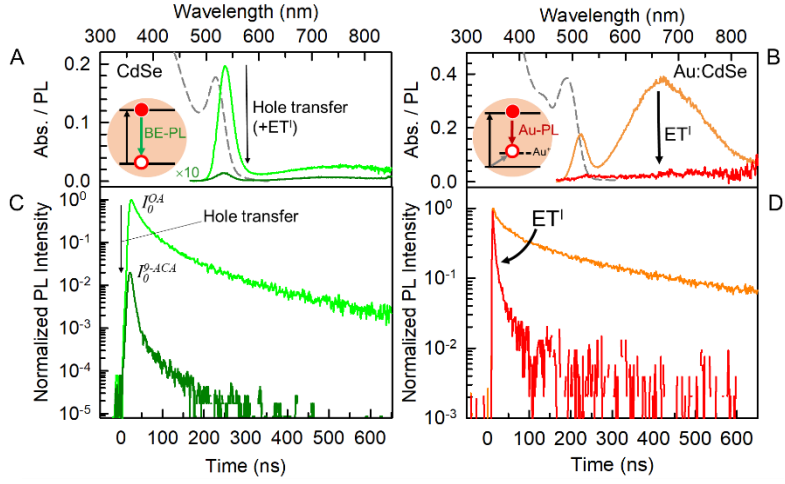

**Figure 2 | Optical properties of CdSe and gold-doped Au:CdSe nanocrystals (NCs).** Photoluminescence (PL) spectra of (**A**) undoped or (**B**) Au-doped CdSe NCs with either OA (light green and orange curves) or 9-ACA (dark green and red curve) ligands in toluene under a 460 nm CW excitation. The optical absorption spectra of the native OA-capped NCs are reported as dashed lines. The insets show depict the emission mechanisms in undoped and Au-doped NCs. Time resolved PL intensity of the same (**C**) undoped or (**D**) Au-doped CdSe NCs with either OA or 9-ACA ligands recorded at the respective PL maximum (550 nm and 670 nm respectively) under pulsed laser excitation at 532 nm. Excitation intensity 0.5 mW cm$^{-2}$, absorbance 0.13 OD at 532 nm for all samples. The data are normalized to the intensity of the respective pristine sample. Measurement conditions are kept unchanged between pristine and 9-ACA functionalized samples so as to enable quantitative comparison between the respective PL intensities and dynamics.

The large Stokes-shift, $\Delta E_{VB-Au} \sim 560$ meV, separating the Au-PL from the respective 1S absorption peak corresponds to the energy difference between the VB and the *t*-states of the Au dopants resulting from the splitting of the *d*-manifold by the crystal-field. The residual band-edge PL observed for the Au:CdSe NCs is due to a minor subpopulation of undoped CdSe NCs in the ensemble. Consistent with the reduced wavefunction spatial overlap between the CB electron and the Au-localized hole with respect to the intrinsic band-edge exciton,[47] the Au-PL in as-sensitized Au:CdSe NCs features a slower kinetics (**Figure 2D** and Fig. S3). Specifically, the decay is multi-exponential with faster components ascribed to non-radiative losses likely associated to trapping in surface defects (notice that neither the doped NCs are passivated with wide bandgap shells) and a slower dominant tail due to the decay of bound excitons. The effective lifetime of the Au-PL extracted as the weighted average between the decay components is $\tau_{Au-PL} \sim 153$ ns.

To estimate the efficiency of the ET′ channel and thereby to assess the effective suppression of non-radiative losses by hole transfer to the TC in Au-doped CdSe NCs, we monitored the relative PL intensity and dynamics of undoped and doped NCs upon replacing the native OA ligands with 9-ACA. As shown in **Figure 2A** and **2B**, in both cases the functionalization with 9-ACA leads to the essentially complete suppression of the *cw* PL. Crucially, however, the corresponding time-resolved PL measurements reported in **Figure 2C** and **2D** – also confirmed by the UC efficiency measurements reported in **Figure 3** – reveal that the photophysical process underpinning the observed PL quenching is drastically different in doped vs. undoped NCs. Specifically, upon replacing OA with 9-ACA, the undoped NCs show a dramatic drop of the zero-delay PL intensity ($I_0^{9-ACA} = \eta \times I_0^{OA}$ with $\eta = 0.02$) accompanied by the substantial acceleration of their PL lifetime (estimated as the time after which the PL intensity has dropped by a factor *e*). Consistent with previous reports [46], the zero-delay drop is ascribed to ultrafast exciton dissociation by hole transfer to 9-ACA – with some minor contribution by hole trapping in surface defects due to incomplete surface coverage by 9-ACA molecules – occurring on a timescale faster than our experimental resolution and affecting the vast majority (>95%) of the NC population. The accelerated PL decay, on the other hand, is caused by ET′ to 9-ACA in the complementary minor fraction (<5%) of the NC ensemble, in which hole transfer is inefficient. From the acceleration of the PL lifetime ($\tau_{BE-PL}^{9-ACA} \sim 5$ ns vs. $\tau_{BE-PL}^{OA} \sim 40$ ns), we estimate that, in such a subpopulation of NCs, $\Phi_{ET'} = 1 - \tau_{BE-PL}^{9-ACA}/\tau_{BE-PL}^{OA} \sim 88\%$. However, because of the ultrafast hole-transfer channel dominating the exciton decay, the global yield for the TC triplet sensitization process in the whole ensemble of undoped NCs is only $\Phi_{ET'}^G = \eta \Phi_{ET'} = 1.8\%$.

A strikingly different behavior is found for the PL decay of Au:CdSe NCs (**Figure 3D**). Specifically, upon replacing OA with 9-ACA, the Au-PL undergoes exclusively a substantial acceleration of its kinetics, with no drop of its zero-delay intensity ($\eta = 1$). This behavior strongly suggests that the hole-routing process in Au:CdSe NCs is effective in preventing the ultrafast hole transfer to the HOMO level of 9-ACA in the *whole* NC ensemble. Consistently, the ET′ process shorten the PL decay time from $\tau_{Au-PL}^{OA} = 153$ ns to $\tau_{Au-PL}^{9-ACA} \sim 4$ ns. As a result, the global triplet sensitization yield coincides with the efficiency of ET′ estimated from the decay kinetics, thus obtaining the remarkable value of $\Phi_{ET'}^{G} = \eta\left(1 - \tau_{Au-PL}^{9-ACA} / \tau_{Au-PL}^{OA}\right) = 97\%$ ~100%, over 50-fold higher than in undoped NCs, which is unprecedented for this class of hybrid sensitizers for sTTA-UC.

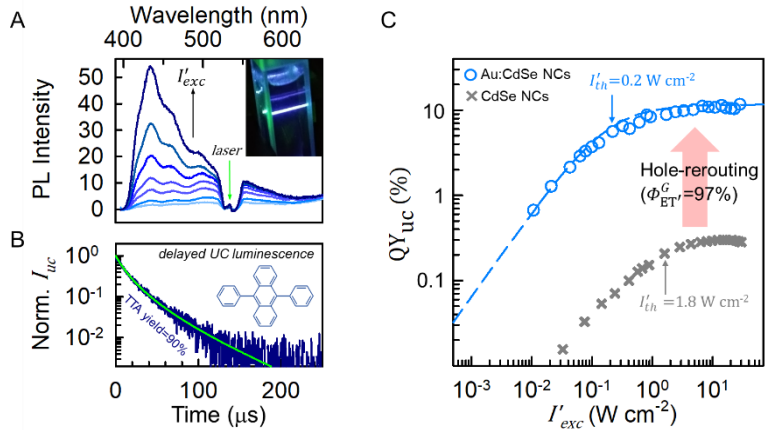

**Figure 3 | sTTA-UC with NC-based hybrid sensitizers. A.** Photoluminescence (PL) spectra of a toluene solution of DPA ($10^{-2}$ M) and Au:CdSe NCs functionalized with 9-ACA (OD 0.13 at 532 nm, optical path 0.1 cm) at increasing 532 nm cw absorbed excitation fluence ($I'_{exc}$). The excitation laser stray light was partially removed with a notch filter. The spectra are normalized with respect to the residual excitation laser intensity. Inset: photograph of the solution under 532 nm excitation showing the upconverted luminescence by DPA molecules. **B.** Time decay of the upconverted DPA-PL intensity ($I_{uc}$) at 440 nm excited by a 532 nm pulsed laser (10 W cm$^{-2}$). The solid line is the fit of the experimental data to the analytical function that describes the time evolution of $I_{uc}$ reported in the main text. **C.** Upconversion quantum yield, QY$_{uc}$, as a function of the absorbed excitation intensity $I'_{exc}$ at 532 nm (bottom axis). The dashed line is the theoretical value of QY$_{uc}$ vs. $I'_{exc}$ calculated as described in the Supplementary Information. The vertical arrows mark the excitation intensity threshold values using doped (blue arrow) or undoped (grey arrow) CdSe NCs as light harvesters.

The effectiveness of the photohole routing pathway in the Au:CdSe NCs in maximizing the sensitization of the 9-ACA triplets is further demonstrated by its exceptional impact on the whole sTTA-UC process, including the final steps of triplet generation and fusion. This was investigated by employing 9,10 dyphenilanthracene (DPA), the mostly studied molecular annihilator/emitter used in sTTA-UC[8,9,11], to realize an upconverting solution in which the global UC yield is determined exclusively by the effectiveness of triplet sensitization process. Through this approach, we could investigate our strategy quantitatively and independently from other parameters that are usually unknown, such as the TTA ability of the annihilator/emitter or its photoluminescence properties. We thus monitored the intensity of the upconverted blue luminescence of DPA under selective excitation of 9-ACA-functionalized Au:CdSe NCs side-by-side to their undoped counterparts. The emission spectra of the upconverting mixture of DPA and the 9-ACA-functionalized Au:CdSe NCs under increasing cw laser excitation fluence at 532 nm (that is, below the energy gap of DPA) are reported in **Figure 4A**, showing the characteristic emission profile of DPA peaked at 440 nm (the spectra for the solution featuring the undoped NCs are shown in Fig. S4). The corresponding time decay curve measured using pulsed 532 nm laser excitation at high intensity (10 W cm$^{-2}$) is shown in **Figure 4B**. Consistent with previous reports, the decay kinetics of the delayed fluorescence intensity ($I_{uc}$) excited via sTTA-UC is orders of magnitude slower than the DPA prompt fluorescence ($\tau \sim 8$ ns)[52], and it is described by the relation $I_{uc}(t) \propto \left(\frac{1-\Phi_{TTA}}{e^{k_T t}-\Phi_{TTA}}\right)^2$, where $k_T$ is the spontaneous decay rate of the DPA triplet and $\Phi_{TTA}$ is the TTA yield.[53,54] The fit of the experimental data yields $k_T = 6.8$ kHz, in agreement with the literature, and $\Phi_{TTA}$ as large as 0.90 (i.e. 90%), which is consistent with the high excitation intensity employed (vide infra).[55,56] Finally, in **Figure 3C** we compare the evolution of the global sTTA-UC quantum yield (QY$_{uc}$, expressed as the ratio between the number of upconverted photons emitted by DPA and the number of incident photons absorbed by the NC sensitizers) with the absorbed excitation intensity ($I'_{exc}$, see Supplementary Material) in upconverting solutions of DPA with either 9-ACA-functionalized undoped or Au-doped CdSe NCs. Both samples behave according to the bimolecular nature of the TTA process in which the triplet annihilation rate ($k_{TTA}$) is proportional to the density of emitter triplets $[T]$ by $k_{TTA} = \gamma_{TT}[T]$, where $\gamma_{TT}$ is the second order rate constant that characterize the TTA.[57] Specifically, in the low excitation regime, where $k_{TTA} \ll k_T$, QY$_{uc}$ (as well as $\Phi_{TTA}$) grows linearly with increasing $I'_{exc}$ and gradually plateaus to its maximum value in the high excitation regime, when the triplet density is so large that TTA becomes the dominant deactivation channel for the emitter triplets. The excitation intensity at which QY$_{uc}$ is half of its saturation value is typically referred to as the excitation threshold for TTA, and is a relevant figure of merit that marks the beginning of the high excitation regime.[46,58] It is worth noting that, as detailed in the Supplementary Material, both QY$_{uc}$ and the excitation threshold $I'_{th}$ depend on the efficiency of the ET′ step by which the NC sensitizes the 9-ACA triplet: the first grows linearly with $\Phi_{ET'}^G$ whereas the latter scales with $\left(\Phi_{ET'}^G\right)^{-1}$. As a result, owing to the effective suppression of non-radiative losses due to hole transfer that boosts $\Phi_{ET'}^G$ from the Au:CdSe NCs to their 9-ACA ligands, the upconversion efficiency saturates at the unprecedented value of $QY_{uc} = 12 \pm 1\%$ ($24 \pm 2\%$ in the *normalized* convention), which is nearly forty-times higher than for the standard undoped NCs ($QY_{uc} = 0.3\%$) in good agreement with the observed enhancement of $\Phi_{ET'}^G$. This result represents the highest upconversion efficiency value reported to date for hybrid sensitizers for *s*TTA, and unambiguously confirms the validity of the proposed exciton manipulation approach to efficiently drive the energy migration across from NCs to conjugated organic ligands. As a further demonstration it should be noted that, as highlighted in **Figure 3C,** through the use of Au-doped NCs the upconversion excitation threshold is lowered by nearly one order of magnitude with respect to the control solution (from 1.8 W cm$^{-2}$ to 0.2 W cm$^{-2}$), thus enabling efficient UC at substantially lower illumination densities thanks to their enhanced triplet sensitization ability.

In summary, we have demonstrated a novel strategy to boost the efficiency of hybrid *s*TTA upconversion systems by using electronic-doped semiconductor NCs as light harvesters to populate the long-lived triplet states of surface-attached conjugated organic moieties. The key aspect of our design is the controlled introduction of a hole-accepting state associated with the electronic dopant within the forbidden gap of the NC that rapidly routes the photohole to an energy above the HOMO level of the organic ligand. This enables us to completely suppress excitation losses by non-radiative hole transfer that is a detrimental parasitic process strongly limiting the efficiency of conventional NC-organic sensitizers for *s*TTA-UC. Owing to their nearly 100% efficient ET yield from the NC exciton to the ligand triplets, our surface functionalized doped-NCs enabled us to obtain photon upconversion yield as high as 12% (24%), representing a record performance for hybrid upconverters based on *s*TTA. We highlight that the strategy proposed here is not limited to the reported material system, but it could be readily applied to any hybrid architecture which requires tuning of the band alignment between its constituents. As such, our approach can be translated to engineer narrower bandgap NCs and obtain efficient, low-power upconverters from the near infrared region that can be directly coupled to existing devices to recover sub bandgap photons, thus further pushing these hybrid nanomaterials towards application in real-world technologies. Indeed, the nanocrystal size, composition, shape and dopant species, as well as the energy-receiving moiety (either organic ligands or other inorganic nanoparticles) can in principle be adapted to rationally design multicomponent systems for upwards as well as for down-conversion photonic applications that require efficient energy transport across any nano-interface without incurring into losses due to ultrafast carrier transfer, including photochemical synthesis, photoredox catalysis, photo-stimulation of biologic and metabolic processes and singlet oxygen generation for photodynamic therapy.

## Methods

*Chemicals*. Gold(III) chloride hydrate, (99.99%)**,** L-glutathione, GSH, (≥98%), tetrabutylammonium bromide, TOABr, (≥98%), 1-dodecanethiol, DDT, (≥98%), ultrapure water (Chromasolv Plus, for HPLC), sodium myristate (≥99%), selenium powder-100 mesh (99.99%), oleic acid, OA (≥90%), cadmium nitrate tetrahydrate (≥98%), 2-propanol (≥99.8%), ethanol (≥99.8%), methanol (≥99.8%), toluene (≥99.5%), 1-octadecene, ODE (≥90%), 9-Anthracenecarboxylic acid, 9-ACA, and 9,10-Diphenylanthracene, DPA, were purchased from Sigma-Aldrich.

*Synthesis of Gold Clusters*. The synthesis of Au$_7$-GSH clusters was carried out following a published procedure[59], mixing 2 mL of HAuCl$_4$ (0.02 M) and 0.6 mL of L-glutathione (GSH, 0.1 M) with 17.4 mL of ultrapure water. After 24 hours of vigorous stirring at 70°C, the clusters were purified by adding isopropanol to the solution (1:2 volume ratio) and centrifuged at 6500 rpm for 20 minutes. This procedure was repeated thrice, and the purified clusters were then dispersed in ultrapure water.

*Ligand exchange on Au clusters*. In 5 mL of Au$_7$-GSH solution (76 μM) were added ~0.3 mL of NaOH 1 M, until pH ~ 9.0, and 5 mL of TOABr, 0.02 M in ethanol. After 2 minutes of vigorous stirring, 5 mL of DDT (0.15 M) in toluene were also added. The mixture was then heated to 70°C and kept under vigorous stirring for 1 hour. The organic phase was separated and washed thrice with ultrapure water to remove water-soluble impurities. Then the Au$_7$-DDT were dispersed in ODE.

*Synthesis of CdSe and Au:CdSe Nanocrystals*. The synthesis of CdSe NCs was carried out following a published procedure.[60] The Cd-myristate precursor was prepared via ex-situ method: a solution of cadmium nitrate in methanol (0.05 M, 40 mL) was added to a solution of sodium myristate in methanol (0.025 M, 240 mL). The white precipitate was washed twice with methanol and dried under vacuum to remove all solvents. 0.1 mmol of Cd-myristate were then added to 0.05 mmol of Se powder in a 25 mL flask with 6.38 mL of ODE and 1 mL of OA and exposed to vacuum for 15 minutes. Successively, the reaction mixture was heated to 210°C for 1 hour under N$_2$ flow. The as obtained NCs were finally purified twice with hexane/ethanol co-solvents by centrifugation (4500 rpm, 10 min) to remove excess ligands and unreacted precursors. The synthesis route for the doped NCs is similar, adding 2.5 mL of Au$_7$-DDT 76 μM to the reaction mixture. The procedure for Au:CdSe with diameter 2.9 nm is the same but the reaction mixture was heated for 2h. In order to synthesize nanocrystals with diameter 2.3 nm, 2 mL of OA were added to the synthesis solution.

*Ligand exchange on NCs*. The procedure was performed in a nitrogen-filled glovebox. 1 mL of 9-ACA in toluene (2.5 mg/mL) was added to 0.5 mL of Au:CdSe dispersed in toluene. The solution was stirred overnight at 50°C. The ligand exchange procedure adopted yields about three molecules per nanocrystal on the surfaces.[36]

*Structural characterization of nanocrystals*. High-Resolution Transmission Electron Microscopy (HR-TEM) imaging was performed on a JEOL JEM-2200FS microscope equipped with a field emission gun working at an accelerating voltage of 200 kV, a CEOS spherical aberration corrector of the objective lens, allowing to reach a spatial resolution of 0.9 Å, and an in-column Omega filter. Powder X-ray Diffraction (XRD) patterns were acquired in Bragg−Brentano geometry with CuKα radiation (Panalytical X'Pert Pro powder diffractometer).

*Photophysical studies.* The solutions absorption spectra were recorded with a Varian Cary 50 UV-Vis spectrophotometer in normal incidence conditions using 1 mm thick glass cuvettes. The steady-state continuous wave (*cw*) photoluminescence (PL) measurements were performed with a Coherent Verdi *cw* laser at 532 nm as excitation source, with a nitrogen-cooled charge-coupled device (CCD Spex 2000) coupled to a polychromator (Triax 190 from J-Horiba) with a spectral resolution of 0.5 nm for signal detection. The excitation laser beam had spot diameter of 140 μm, containing 90% of intensity, with Gaussian profile. Shape and spot size were measured according to the *knife-edge* method. The source irradiance was measured by means of a Thorlabs S120VC photodiode power sensor, so the power densities were calculated from the incident power and spot size. $^{-5}$ M in ethanol, QY=95%). For the upconverting samples, the maximum yield was measured under a CW 532 nm excitation source in normal incidence conditions, using glass cuvettes of 1 mm optical path at the excitation intensity corresponding to the maximum PL intensity.

The time-resolved PL measurements were carried out using a 532 nm II harmonic of a Nd:YAG pulsed laser (Laser-Export Co. LSC-DTL-374QT, pulse width 5 ns), whereas the UC signal was recorded modulating the CW 532 nm laser with a TTi TG5011 wavefunction generator, with a time resolution better than 0.1 μs. The emission wavelength was selected by Cornerstone monochromator (15 nm band pass) and the signal intensity was recorded using a Hamamatsu R943-02 photomultiplier coupled with an Ortec 9353 multichannel scaler, with a temporal resolution of 0.1 ns. The excitation intensity was modulated by means of neutral filters with varying optical density (Thorlabs) and the laser stray light was reduced using a 532 nm notch filter. All the collected spectra were corrected for the instrumental spectral response. The photoluminescence quantum yield of Au-doped nanocrystals ($2.68 \times 10^{-5}$ M solution in toluene) and of the upconversion systems was measured by relative methods using Rhodamine 6G as standard reference (10-

**Acknowledgements**
Financial support from the Italian Ministry of University and Research (MIUR) through grant "Dipartimenti di Eccellenza - 2017 Materials For Energy" is gratefully acknowledged.

**Author contributions**
A.M and S.B. conceived the study. A.R., A.M. and F.M. designed and performed the upconversion spectroscopy measurements, analysed the data and elaborate the model of *s*TTA in upconversion in hybrid systems. S.B., G.G. and C.C. designed and realized the nanocrystals. A.R., V.P. and M.L.Z. performed spectroscopy experiments and analysed the data. A.M. and S.B. wrote the manuscript in collaboration with all authors.

**Additional information**
Correspondence and requests for additional information materials should be addressed to S.B. and A.M..

**Competing financial interests**
The authors declare no competing financial interests.